\documentclass[12pt]{article}
\usepackage{epsfig}
\begin{document}
\setlength{\baselineskip}{0.30in}
\newcommand{\nc}{\newcommand}
\newcommand{\beq}{\begin{equation}}
\newcommand{\eeq}{\end{equation}}
\newcommand{\be}{\begin{eqnarray}}
\newcommand{\ee}{\end{eqnarray}}
\newcommand{\num}{\nu_\mu}
\newcommand{\nue}{\nu_e}
\newcommand{\nut}{\nu_\tau}
\newcommand{\nus}{\nu_s}
\newcommand{\mnus}{M_s}
\newcommand{\taus}{\tau_{\nu_s}}
\newcommand{\nnt}{n_{\nu_\tau}}
\newcommand{\rnt}{\rho_{\nu_\tau}}
\newcommand{\mnt}{m_{\nu_\tau}}
\newcommand{\tnt}{\tau_{\nu_\tau}}
\newcommand{\bi}{\bibitem}
\newcommand{\rar}{\rightarrow}
\newcommand{\lar}{\leftarrow}
\newcommand{\lrar}{\leftrightarrow}
\newcommand{\dm}{\delta m^2}
\newcommand{\so}{\, \mbox{sin}\Omega}
\newcommand{\co}{\, \mbox{cos}\Omega}
\newcommand{\sotil}{\, \mbox{sin}\tilde\Omega}
\newcommand{\cotil}{\, \mbox{cos}\tilde\Omega}
\makeatletter
\def\alt{\mathrel{\mathpalette\vereq<}}
\def\vereq#1#2{\lower3pt\vbox{\baselineskip1.5pt \lineskip1.5pt
\ialign{$\m@th#1\hfill##\hfil$\crcr#2\crcr\sim\crcr}}}
\def\agt{\mathrel{\mathpalette\vereq>}}

\newcommand{\eq}{{\rm eq}}
\newcommand{\tot}{{\rm tot}}
\newcommand{\M}{{\rm M}}
\newcommand{\coll}{{\rm coll}}
\newcommand{\ann}{{\rm ann}}
\makeatother

\begin{center}
\vglue .06in
{\Large \bf {Massive sterile neutrinos as warm Dark Matter}}
\bigskip
\\{\bf A.D. Dolgov
\footnote{Also: ITEP, Bol. Cheremushkinskaya 25, Moscow 113259, Russia.}
\footnote{e-mail: {\tt dolgov@fe.infn.it}},
S.H. Hansen\footnote{e-mail: {\tt sthansen@fe.infn.it}}
 \\[.05in]
{\it{INFN section of Ferrara\\
Via del Paradiso 12,
44100 Ferrara, Italy}
}}
\\[.40in]
\end{center}

\begin{abstract}
We show that massive sterile neutrinos mixed with the ordinary ones may 
be produced in the early universe in the right amount to be natural warm 
dark matter particles. Their mass should be below 40 keV 
and the corresponding mixing angles sin$^2 2 \theta > 10^{-11}$
for mixing with $\num$ or $\nut$, while mixing with $\nu_e$ is 
slightly stronger bounded with mass less than $30$ keV.
\end{abstract}    


\section{Introduction}

There seems to be convincing experimental evidence for non-zero neutrino 
masses and mixing angles (for a review see e.g. \cite{rev}), and if all 
the present day data are correct, there must exist at least one sterile 
neutrino species. These neutrinos should be very light (sub eV range) and 
hence contribute negligibly to the cosmological energy 
density~\cite{gerstein}
\be
\Omega_\nu h^2 = \frac{m_\nu}{92 \mbox{eV}} ~\ ,
\ee
if they were produced with the equilibrium number density in the early 
universe at a temperature 
below $\sim 10$ MeV. If their number density was smaller
than the equilibrium one, then the permitted value of the mass could be
correspondingly higher.

One could easily envisage more than one sterile neutrino species. The
masses and mixing angles of these extra neutrinos are essentially free
parameters.  If we consider sterile neutrinos with masses $10-200$
MeV, then big bang nucleosynthesis and energy loss arguments for SN
1987A allow one to exclude mixing angles in the range sin$^2 2 \theta
= 10^{-1} - 10^{-12}$~\cite{dolgov00}, where both the upper and lower
limits vary as functions of mass. On the other hand, direct
terrestrial experiments exclude supplementary mixing angles in the
range sin$^2 2 \theta = 0.001 - 1$~\cite{nomad}, and the lower limit
is weaker for small masses, $m \sim 10$ MeV.
We should mention that the excluded regions only overlap for some
masses, and there is still some non-excluded parameter space for
$m<40$ MeV and sin$^2 2 \theta = 0.01 - 1$.

The hypothesis that sterile neutrinos could make a
considerable contribution to cosmological dark matter has a rather
long history. The idea that right-handed sterile
neutrinos may form warm dark matter was briefly discussed in
ref.~\cite{olive82} and was further pursued in the 
paper~\cite{dodelson94}. In more detail warm dark matter cosmology 
was considered in ref.~\cite{colombi96}. 
There are some other warm dark matter candidates discussed in the
papers~\cite{malaney95,dvali98}. Sterile neutrinos coming from
a mirror world may also make WDM in our universe; they were discussed
in the papers~\cite{berezhiani}. More models and references can be
found in the recent works~\cite{larsen99,hannestad00}.
A dark matter model with sterile neutrinos but with a non-thermal
spectrum was considered in ref.~\cite{shi99}. Such neutrinos could be
produced by the resonance oscillations in the early universe in the
presence of a large lepton asymmetry. This model was further considered
in ref.~\cite{drees00}, where constraints originating from 
consideration of decays of $\nus$, especially of the radiative one, 
were presented.
A detailed analysis of the essential physics of this model, namely
the importance of the mixing angle suppression in the early universe, is made
in ref.~\cite{fuller}.

In this paper we find the allowed values of 
mass and mixing angle of a sterile 
neutrino, $\nu_s$, so that the latter could be a dominant dark matter 
particle. We will consider the production of $\nu_s$
in the early universe, calculate their energy spectrum,
discuss their different decay modes, and derive bounds on 
mass and mixing angle from cosmology and astrophysics.

\section{Production of $\nus$ in the early universe}

Let us consider for simplicity a two-neutrino mixing scheme, 
where one of the active neutrinos, $\nu_a = \nue, \num$ or $\nut$, 
mixes with a heavy mainly sterile neutrino, $\nus$,
\be
\nu_a &=& \cos\theta ~\nu_1 + \sin \theta ~\nu_2\,,  \nonumber \\
\nu_s &=& -\sin \theta ~\nu_1 + \cos\theta ~\nu_2\,,
\label{nuas}
\ee
where $\nu_1$ and $\nu_2$ are assumed to be the light and heavy mass
eigenstates respectively, and $\theta$ is the vacuum mixing angle. 
We will consider small mixing angles, and hence sometimes refer to the 
light neutrino mass eigenstate as the active neutrino and the heavy one 
as sterile neutrino.

We assume that sterile neutrinos were initially absent in the primeval 
plasma and were produced through the mixing with active ones.
As will be clear shortly, the temperature suppression of the effective
mixing angle changes the production rate at high temperatures from $T^3$
to $T^{-9}$, so with small mixing angles the sterile neutrinos would never
have reached equilibrium. As was estimated in ref.~\cite{barbieri}
the maximum ratio of the production rate to the expansion rate takes 
place at rather low temperatures, $T \approx 0.1 \mbox{GeV}
(m/\mbox{MeV})^{1/3}$.
The production rate is usually approximated as~\cite{barbieri,enqvist}
\be 
\Gamma/H = \frac{\mbox{sin}^2 2 \theta_M}{2} \left( \frac{T}{T_W}
\right)^3 \, ,
\label{scat}
\ee
where H is the Hubble expansion parameter, $T$ is the plasma
temperature, and $T_W$ is the decoupling temperature of the active 
neutrinos, which is approximately taken about $T_W = 3$ MeV. 
Instead of this approximate equation, below
we will write down and solve the exact momentum dependent Boltzmann 
equation, taking into account the processes of production of 
$\nus$ but neglecting inverse reactions. The latter are not important
if the number density of $\nus$ is small. In this more precise
approach the question of the value of $T_W$ never appears, it is
solved automatically. Another advantage of our approach here is
that it permits to calculate the energy spectrum of $\nus$,
while the previous method permitted only to estimate the total 
number density.
Before doing these calculations it may be instructive to make the
standard simplified estimates and later compare them with the 
exact results found below. 

The mixing angle sin$^2 2 \theta_M$ is suppressed at large temperatures 
due to matter effects, and for $\num$ or $\nut$ mixing
it can be written as~\cite{notzold88}
\be
\sin 2 \theta_\M \approx {\sin 2\theta \over
1+ 0.8 \times 10^{-19}\, (T_\gamma/\mbox{MeV})^6 
(\delta m^2/\mbox{MeV}^2)^{-1}}~,
\label{thetam}
\ee
where the coefficient in front of the second term in the denominator
was obtained by a rather arbitrary procedure of
thermal averaging of the factor 
$\langle E^2 \rangle \approx 12 T^2$, entering the ratio of the
neutrino refraction index to the vacuum term $\delta m^2 /2 E^2$.
We see from this expression that matter effects become essential 
and suppress the mixing for 
$T_\gamma>0.15\,\mbox{GeV}\,  (m/\mbox{keV})^{1/3}  $ (a
similar argument was made in
ref.~\cite{dodelson94} for the left-right neutrino mixing). 
For the $(\nu_e-\nus)$-mixing the
factor in the denominator should be $2.7$ instead of $0.8$.
We will assume here that $\dm = m_2^2 - m_1^2$ is positive 
(specifically we assume $m_1 \ll m_2$), and if instead
the active neutrino is heavy the analysis somewhat changes~(see 
refs.~\cite{shi99,fuller}).

For the energy dependent calculations we need the expression
for the matter effects in the denominator of eq.~(\ref{thetam}) 
prior to averaging over the thermal bath. The latter can be read off
from the relevant equations of refs.~\cite{barbieri,notzold88}
\be
\sin 2 \theta _M = \frac{\sin 2 \theta}{1+ 3.73 \cdot 10^{-20}\,
c_2\,m({\rm MeV})^{-2}\,( y^2/x^6) } \, ,
\label{supp}
\ee
where the $\nus$ mass, $m$, is measured in MeV and 
we used the expansion parameter of the universe, $a$, to introduce 
the new variables, $x = 1 {\rm MeV} \times a$,
$y = E a$, and neglected a possible entropy release so that the
temperature drops according to $T= 1/a$. The numerical coefficient
$c_2$ depends upon the neutrino flavour: $c_2=0.61$ for $\nue$ and
$c_2 = 0.17$ for $\nut$ and $\num$. However, for the temperatures
close to or above the muon mass $c_2$ becomes the same for $\nue$ 
and $\num$.

The Boltzmann equation describing the evolution of the sterile neutrino
distribution function, $f_s$, in terms of these new variables takes the
form
\be
xH \partial _x f_s = I_{\rm coll} \, ,
\label{kineq}
\ee
where the collision integral is given by
\be
I_{\rm coll} =  \frac{1}{2E_s} 
\int \frac{d^3 p_2}{(2\pi)^3 2E_2} |A|^2 f_3 f_4 d\tau_{3,4}  \, ,
\ee
where $ d\tau_{3,4}$ is the phase space element (together with
the energy-momentum $\delta$-function) of the particles $l_3$ and $l_4$
in whose collision $\nus$ is produced, 
\be
l_3 + l_4  \rightarrow \nu_s + l_2 \, ,
\ee
and $f_{3,4}$ are their
distribution functions. We assume that the latter are equal to
their equilibrium values and then the
conservation of energy gives $f_3 f_4 = \exp (-y_1-y_2)$
in the Boltzmann approximation.
Integrating the probabilities of 
all the relevant processes over phase space (see the appendix)
allows to find the collision integral and 
to solve equation~(\ref{kineq}) analytically, giving
\be
f_s = 3.6 \cdot 10^8\,
 \sin^2 \theta  \, \left( 1 + g_L^2 + g_R^2 \right)
 \, c_2^{-1/2} \, m({\rm MeV}) 
\, \left(\frac{10.75}{g_*}\right)^{1/2} f_a \, ,
\label{dmprod}
\ee
where $f_a $ is the distribution function of 
any of the active neutrinos, and $g_*$ is the number
of relativistic degrees of freedom at the time when the sterile neutrinos
were produced.  Subsequent to the
production there will be a dilution of the active neutrinos relative to 
the sterile ones. This is described by another factor $(g_*/10.75)$. 

The coefficient relating $f_s$ to $f_a$ in eq.~(\ref{dmprod}) is
independent on the energy of neutrinos, so the spectrum of $\nus$
remains the same as that of active neutrinos. This is somewhat surprising 
because the reaction rate is proportional to the neutrino energy. However,
for smaller $E$ the rate becomes efficient at higher temperature, as 
one can see from the expression (\ref{supp}) describing the suppression
of neutrino mixing in matter. This effect compensates the factor of 
momentum, $y$,
in the kinetic equation. However, at very large temperatures,
$T\sim M_{W,Z}\sim 100$~GeV, the weak reaction rate drops down, and hence
the spectrum will be somewhat distorted at very small $y$'s.

It is interesting to compare the accurate results presented above with
the simplified calculations based on the solution of the following
approximate kinetic equation
\be
Hx \, \partial _x f_s = 
\frac{1}{2} \sin^2 2\theta_M \, \Gamma_W \, f_a \, ,
\ee
where the mixing angle and interaction rate can be taken from 
eqs.~(\ref{scat}, \ref{thetam}). This equation is easily integrated, 
and we find that the result for the total number density of $\nus$
agrees within a factor of 2 with the more accurate
result~(\ref{dmprod}).

Up to now we have seen how the produced amount of sterile neutrinos depends on 
the mass and mixing angle, so let us
instead ask: how many sterile neutrinos should be produced in order for
them to be a dark matter candidate? Let us take $\Omega_{\rm DM}  =0.3$,
which means that we must demand $\rho_s = 3 \, h^2 \, {\rm keV}/{\rm cm}^3$.
Using $h=0.65$ and $n_\alpha^{\rm today} = 100/{\rm cm}^3$ one finds
\be
n_s = 1.27 \times 10^{-5} \, n_\alpha \, \left(\frac{\mbox{MeV}}{m}\right)
\left( {\Omega_{DM} \over 0.3}\right)
\left( {h \over 0.65}\right)^2 \, ,
\label{dmneed}
\ee
and comparing eqs.~(\ref{dmprod}) and ~(\ref{dmneed}) one obtains
\be
\sin ^2 \theta = 3.6 \times 10^{-14} \, \frac{c_2^{1/2}}{
\left( 1 + g_L^2 + g_R^2 \right)} \left( \frac{10.75}{g_*}\right) ^{1/2} 
 \left( \frac{\mbox{MeV}}{m} \right)^2  \, .
\label{comp}
\ee

This equation thus describes a line in mass-mixing parameter space, where
the sterile neutrino must lie, if it indeed is the dominant dark matter
particle. Let us now see how decay processes and supernovae 
can further restrict this parameter space.

\section{Decay}
The mixing couples the heavier $\nu_2$ to the Z-boson, and allows for
the decay channel
\be
\nu_2 \rightarrow \nu_1 + \ell + \bar \ell \, ,
\label{dec}
\ee
where $\nu_1$ is mostly an active flavour and $\ell$ is any lepton with
a mass smaller than half the mass of the heavy neutrino.  
This mixing angle can be translated into decay time
\be
\tau = \frac{10^{5} \, f(m) }{m(\mbox{MeV})^5 \, 
\mbox{sin}^2 2 \theta}\, \mbox{sec} \, ,
\label{decaytime}
\ee
where $f(m)$ takes into account the open 
decay channels (for $m<1$ MeV only the neutrino channels are open, and 
$f(m)=0.86$, while for $m_s >2m_e$ the $e^+e^-$-channel is also open and 
$f(m)=1$). Now, for the sterile neutrino to be a dark matter candidate 
we must demand that it does not decay on cosmic time scales, 
which means $\tau > 4 \times 10^{17}$ sec, and hence from 
eq.~(\ref{decaytime}) we get
\be
\mbox{sin}^2 2 \theta < 2.5 \times 10^{-13} \, 
\frac{f(m)}{m(\mbox{MeV})^5} \, .
\label{time}
\ee

We can, however, get an even stronger bound by considering the radiative 
decay
\be
\nu_s \rightarrow \nu_a + \gamma \, ,
\label{nng}
\ee
where $\nu_a$ is any of the active neutrinos. This decay 
will contribute with a distinct line into the diffuse photon 
background near $m/2$. The branching ratio for
the reaction~(\ref{nng}) was found~\cite{sarkar} to be: $BR \approx 1/128$.
The flux of electromagnetic radiation from the decay was calculated in 
the papers~\cite{stecker80,kimble81} (see also 
refs.~\cite{drees00,kolbturner}). In the case of a large life-time,
larger than the universe age, and of the matter dominated flat universe
the intensity of the radiation in the frequency interval $d\omega$ 
is equal to
\be
dI = (BR)\,{n_s^{(0)} \over H \tau_s}\, {\omega^{1/2} d\omega \over
(m_s/2)^{3/2}}
\label{didom}
\ee
where $n_s^{(0)}$ is the present day number density of $\nus$ and $H$ is
the Hubble constant. Here we neglected corrections related to
a possible dominance of the lambda-term in the latest history of the 
universe.

In the energy range interesting to us a
rather conservative upper limit for the flux can be read of
the figure of ref.~\cite{Ressell:1989rz}
\be
\frac{d {\cal F}}{d \Omega} < 0.1 \, \left( \frac{1 {\rm MeV}}{E} \right)
{\rm cm }^{-2}{\rm sr }^{-1}{\rm sec }^{-1}
\label{dfdo}
\ee
and taking the values
$\Omega_s = 0.3$  and $h=0.65$ we find:
$\tau > 4 \times 10^{22}$, which together with eq.~(\ref{decaytime})
leads to the bound
\be
\mbox{sin}^2 2 \theta < 2.5 \times 10^{-18} 
\, \frac{f(m)}{m(\mbox{MeV})^5} \, .
\label{time2}
\ee

A mass dependent bound can be found by considering the energy 
loss argument for SN 1987A, for $m_s < 3T_{SN} \approx 100$ MeV.
Sterile neutrinos produced due to mixing with the active ones inside the
supernova would carry away too
much energy, hence shortening the explosion. 
The excluded mixing angles have been calculated
several times for SN 1987A, and the results are about sin$^2 2 \theta <
3 \times 10^{-8}$ for $\num$ or $\nut$ mixing~\cite{dolgov00}
and sin$^2 2 \theta < 10^{-10}$ for the $\nu_e$ mixing~\cite{sn} for
masses of the order MeV. For smaller masses, $m < 40$ keV, this bound
weakens substantially due to matter effects, since the
matter mixing angle can be expressed as~\cite{raffelt}
\be
\mbox{sin}^2 2\theta _M = \frac{\mbox{sin}^2 2 \theta}{\mbox{sin}^2 2 \theta +
\left( \mbox{cos} 2 \theta + 1.4 \cdot 10^3 \left( 
\frac{{\rm keV}^2}{m^2} \right)  \right)^2}~.
\ee
On fig.~1 one sees, that this SN-bound is weaker than the diffuse gamma 
background limit for all neutrino masses.

\begin{figure}[htb]
\begin{center}
\epsfig{file=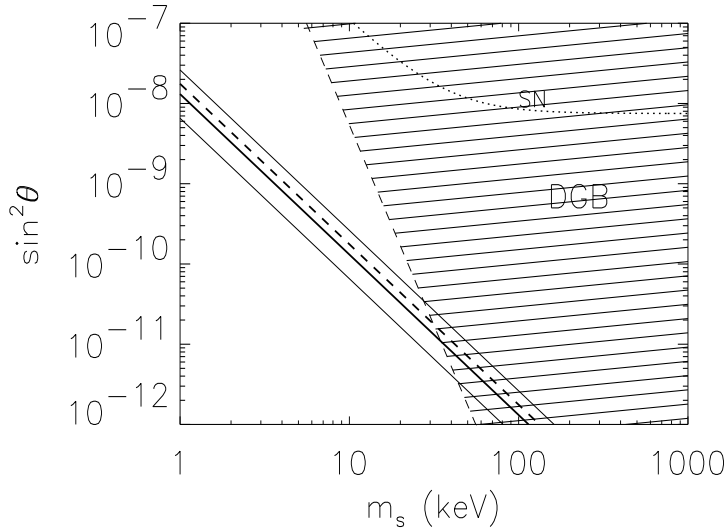,height=8cm,width=10cm}
\end{center}
\caption{
Bounds from $(\nu_\alpha - \nu_s)$-mixing. The middle full line describes 
the mass-mixing relationship if sterile neutrinos are the dark matter for
$(\nu_\tau - \nu_s)$-mixing.
The two other full lines allow a factor 2 uncertainty in the amount
of dark matter, $\Omega_{DM} = 0.15 -0.6$.
The dashed line is for $(\nu_e - \nu_s)$-mixing.
The hatched region for big masses is excluded by the Diffuse Gamma Background.
The region above the dotted line is excluded by the duration of SN 1987A
for $(\nu_\tau - \nu_s)$-mixing.}
\label{fig1}
\end{figure}

Plotting the equation describing the production, eq.~(\ref{comp}),
together with the bounds from SN 1987A and radiative decay, makes it clear
that $(\nu_\alpha-\nu_s)$-mixing (see fig.~1) as the producer of dark matter
demands that the neutrino mass must be smaller than $40$ keV for $\nu_\mu$
or $\nu_\tau$ mixing (solid lines), and slightly below this values for 
$\nu_e$ mixing (dashed line), and with
corresponding mixing angle sin$^2 \theta > 10^{-11}$.
The two thinner full lines in the figure allow for a factor
2 uncertainty in the amount of dark matter, $\Omega_{DM} = 0.15 - 0.60$.

A lower bound on the dark matter particle can be obtained from large
scale structure. A small mass of the dark matter particle will erase
structure on small scale, and present day data seem to exclude masses 
smaller than a few tens of eV.

Several comments are in order here. First we must check that the 
sterile neutrinos are indeed relativistic when produced. 
This is the case, because the temperature $T_{{\rm max}}$ is
about $1.3$ GeV for $m=1$MeV, and about $0.13$ GeV for $m=1$keV.
Further, the dilution factor is somewhere between 1 and 4 depending upon
whether the production happens before or after the QCD transition,
and can thus enlarge the allowed region slightly compared to the
figures, where we for simplicity used $g_*=10.75$.
Looking at eq.~(\ref{dmprod}) it seems that the sterile neutrinos 
follow an equilibrium distribution function. This is not quite the case,
because the small momentum neutrinos are produced first, and hence
their relative importance is increased by the subsequent entropy release
(which dilutes the active neutrinos). A different non-thermal effect can
appear for $(\num-\nu_s)$-mixing, since the factor $c_2$ is $0.17$ when
the $\mu$'s are absent (for $T \ll m_\mu$), whereas it grows to $c_2 = 0.61$
when the muons are fully present in the plasma. This means that bigger
momenta will be produced with a factor $1.9$ more efficiently~\cite{names}.

\section{Discussion and conclusion}
\label{concl}

The model considered in this paper, is undoubtly the simplest, oldest
and, as we have seen, very natural for warm dark matter. The value of
the $\nus$ mass can in the future be found from the detailed analysis
of large scale structure formation.  As we have seen this model
permits masses about keV which may be interesting for the galaxy
formation problem~\cite{larsen99}, and furthermore recent N-body
simulations of large scale structure compared with observations of the
number of satellite galaxies have in fact indicated that the dark
matter particle may have a mass about keV~\cite{nbody}. It is
interesting to note, that a keV sterile neutrino has been suggested as
a possible explanation of the observed pulsar
velocities~\cite{kusenko}. Finally, better observations of the diffuse
$\gamma$ background around keV energies should be able to cut away
more of the parameter space, or potentially make an indirect
observation of dark matter.

\section*{Acknowledgement} It is a pleasure to thank K.~Abazajian, 
Z.~Berezhiani and G.~Fuller for interesting and fruitful conversations, 
in particular SH thanks G.~Fuller for presenting ref.~\cite{fuller}
prior to publication.
AD is grateful to the Theory Division at CERN for hospitality while this
work was done.

\appendix

\section{Solving the Boltzmann equation}
All the relevant processes were presented in table~2 of ref.~\cite{dolgov00}.
There are two kinds of matrix elements, namely $(p_1 p_2)(p_3 p_4)$ and
$(p_1 p_4)(p_3 p_2)$, and one finds from the integral over phase space that
\be
\int d\tau_{34} (p_1 p_2)(p_3 p_4) &=& 
3 \, \int d\tau_{34} (p_1 p_4)(p_3 p_2) = \frac{(p_1 p_2)^2}{8 \pi} \, .
\ee
Now one can count all the relevant processes, integrate over momenta and find
\be
Hx \partial _x f_s = \frac{5 \times 2^4}{3 \pi^3} \, \sin^2 \theta \,
\left( 1 + g_L^2 + g_R^2 \right) \, G_F^2 \, E_1 T^4 \, f_a \, ,
\ee
where $Hx=4.5 \times 10^{-22} \left( \frac{g_*}{10.75}  
\right) x^{-1} $ MeV and $G_F = 1.1664 \times 10^{-5} $
GeV. With the variable $ \xi = y/x^3$ and $\beta$ defined in eq.~(\ref{supp})
the suppression of mixing angle is
\be
\sin 2\theta _M = \frac{\sin 2 \theta}{1+\beta^2 \xi ^2} \, ,
\ee
and with the integral
\be
\int _0 ^\infty d\xi \, \left( \frac{1}{1+\beta^2 \xi ^2} \right) ^2 =
\frac{4}{\pi} \, \frac{1}{\beta} \, ,
\ee
we find the result in eq.~(\ref{dmprod}).

\end{document}